\documentclass{article}
\usepackage{spconf,amsmath,graphicx, amssymb}
\usepackage{subcaption}
\usepackage{bm}
\usepackage{multirow}
\usepackage{booktabs}
\usepackage{tikz}
\usepackage{float}

\title{Breaking the trade-off in personalized speech enhancement with cross-task knowledge distillation}
\name{\begin{tabular}{c}Hassan Taherian$^{1,2*}$, Sefik Emre Eskimez$^{1}$, {\normalfont{and}} Takuya Yoshioka$^{1}$\thanks{*This work was carried out during the internship at Microsoft Research.}\end{tabular}}

\address{$^1$Microsoft, Redmond, WA, USA~~~~~~~~~~~~
$^2$The Ohio State University, Columbus, OH, USA\\{\small \texttt{taherian.1@osu.edu, \{seeskime, tayoshio\}@microsoft.com}}}

\begin{document}
\ninept
\maketitle

\begin{abstract}

% Personalized speech enhancement (PSE) models achieve promising results due to their ability to remove both interfering speech and background noise.
% Unlike the unconditional speech enhancement, causal PSE models may remove the target speech by mistake. The PSE models also tend to leak interfering speech when the target speaker is silent for an extended period. We show that existing PSE methods suffer from a trade-off between the speech over-suppression and the interference leakage by addressing one problem at the expense of the other. We propose a new PSE model training framework using cross-task knowledge distillation to mitigate this trade-off. We utilize a personalized voice activity detector during training to exclude the non-target speech frames that are wrongly identified as the target speaker. This prevents the PSE model from being too aggressive while suppressing the interference leakage. Comprehensive evaluation results are presented, covering various PSE usage scenarios. 

Personalized speech enhancement (PSE) models achieve promising results compared with unconditional speech enhancement models due to their ability to remove interfering speech in addition to background noise. Unlike unconditional speech enhancement, causal PSE models may occasionally remove the target speech by mistake. The PSE models also tend to leak interfering speech when the target speaker is silent for an extended period. We show that existing PSE methods suffer from a trade-off between speech over-suppression and interference leakage by addressing one problem at the expense of the other. We propose a new PSE model training framework using cross-task knowledge distillation to mitigate this trade-off. Specifically, we utilize a personalized voice activity detector (pVAD) during training to exclude the non-target speech frames that are wrongly identified as containing the target speaker with hard or soft classification. This prevents the PSE model from being too aggressive while still allowing the model to learn to suppress the input speech when it is likely to be spoken by interfering speakers. Comprehensive evaluation results are presented, covering various PSE usage scenarios. 
%xperimental results show that our method reduces the interference leakage by more than 100 dB without increasing the speech over-suppression.

\end{abstract}
\section{Introduction}
\label{sec:intro}

Remote meetings have become part of our daily lives in the rapidly emerging hybrid work era. 
Causal and real-time speech enhancement (SE) algorithms are now integrated into most teleconferencing services to attenuate background noise. Meanwhile, personalized speech enhancement (PSE) is gaining increased attention from the research community. PSE utilizes additional cues such as a speaker embedding vector of a target speaker to enhance only the speaker's signal even when interfering speech and background noise are both present~\cite{scPSE, giri2021personalized, taherian2022one}. The PSE task may be regarded as a combination of speech separation, enhancement, and speaker verification tasks.

Despite the advantage over SE, the current causal PSE methods face two major problems, i.e., speech over-suppression and interference leakage. 
Speech over-suppression refers to the problem of the target speaker's voice being identified as an interfering speaker and wrongly removed. This problem is worse for the same-gender mixtures due to voice characteristic similarities between the target and interfering speakers~\cite{delcroix2020improving}. As reported by prior studies, speech over-suppression negatively impacts automatic speech recognition (ASR) accuracy~\cite{Wang2020} and human communication experiences~\cite{scPSE}. 

The second problem, or interference leakage, means that  the PSE models often fail to remove interfering speakers when the target speaker is not present at all or for a sustained period. This problem has yet to be fully investigated, as most prior works assumed the target speaker to be actively speaking. In practical scenarios such as video conferencing, the target speaker can be inactive or silent for a long time.  A naive solution for reducing the interference leakage would be to add inactive target speaker (ITS) samples in the training data~\cite{delcroix2022listen} and train the PSE model to generate zero signals for the ITS samples. However, precisely identifying ITS frames is  challenging due to the causality constraint and the model size limitation. Forcing the PSE model to generate zero signals for all the ITS frames regardless of their difficulty levels results in increased speech over-suppression. 
Previously proposed PSE models suffered from a trade-off between the speech over-suppression and the interference leakage by addressing only one problem at the expense of the other.

%proposed method

We propose a cross-task knowledge distillation approach to reduce both speech over-suppression and interference leakage and thus overcome the trade-off between these two problems. Specifically, we utilize a causal personalized voice activity detector (pVAD) to identify the frames in the ITS training samples that are wrongly classified as the target speaker (note that the ITS samples contain no target speakers). We then modify the PSE loss function based on the pVAD outputs to adjust the contribution of each frame. With the modified PSE loss, we exclude or de-emphasize the misclassified frames as these frames  are  difficult to handle, and including them during training can exacerbate the speech over-suppression. We show the effectiveness of our proposed training method in different scenarios.

% Without increasing the speech over-suppression, the propose method 
% method reduces the interference leakage by more than 100 dB when target speaker is inactive. 

\begin{figure*}[t]
  
  \includegraphics[width=0.95\textwidth]{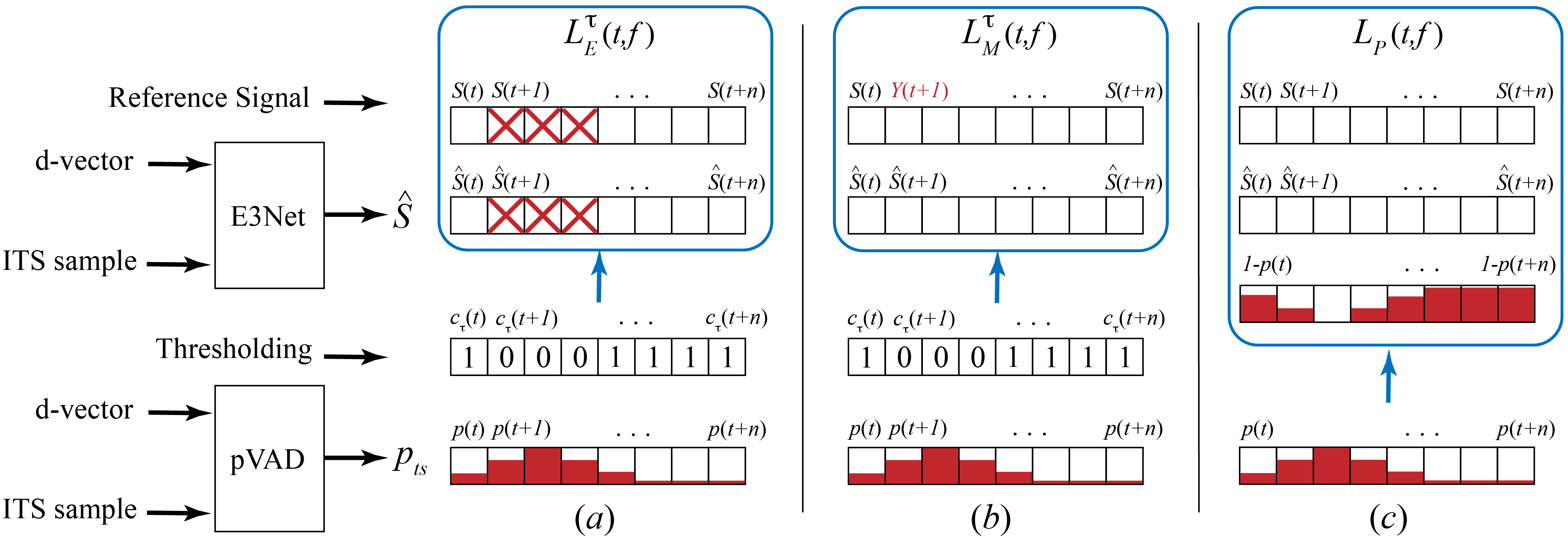}
  \centering
  \caption{Schematic diagram of E3Net training with cross-task knowledge distillation. (a) Misclassified frames are excluded from PSE loss. (b) Noisy signal $Y$ is used as the reference signal for misclassified frames. (c) Active target speaker probabilities are used as weights in PSE loss. In (a), crossed-out frames are excluded from the loss computation.}
  \label{fig_e3net_pvad}
  \end{figure*}

\section{Related Work}

Several studies developed causal PSE models utilizing a speaker embedding vector to extract the target speaker's voice.~\cite{scPSE, giri2021personalized,taherian2022one, thakker2022fast, Chen2022MultiStage}. Giri et al. proposed a perceptually motivated PSE model with low complexity~\cite{giri2021personalized}. In~\cite{scPSE}, two real-time PSE models were proposed and evaluated in various scenarios. Thakker et al. introduced an efficient real-time PSE model with low computational cost~\cite{thakker2022fast}. \cite{Chen2022MultiStage} employed a multi-stage and multi-loss framework to train a full band PSE. In~\cite{9747765}, a dual-stage PSE network is proposed where the target speech magnitude is estimated in the first stage, and the clean phase information is retrieved in the second stage. These studies paid limited attention to the trade-off problem between speech over-suppression and interference leakage.

Wang et al. proposed an asymmetric loss~\cite{Wang2020} for a target speaker extraction system for speech recognition to mitigate the speech over-suppression. 
The asymmetric loss penalizes the time-frequency bins where the target speaker's voice is over-suppressed. While it reduces speech over-suppression, the asymmetric loss significantly increases the interference leakage. \cite{scPSE} proposed a PSE model with ASR-based multi-task training to alleviate the speech over-suppression problem. 

A few recent studies attempted to address the interference leakage problem to handle the case where the target speaker is inactive~\cite{delcroix2022listen, borsdorf21, Zhang2020}. \cite{Zhang2020} and~\cite{borsdorf21} proposed time-domain speaker extraction models with a modified signal-to-noise (SNR) ratio loss for the inactive target speaker scenario. 
\cite{delcroix2022listen} trained a target speaker extraction model with ITS samples using a modified SNR loss that preserves the input signal amplitude at the system's output. To reduce interference leakage, they utilized an extra speaker verification module to detect if the extracted speech belonged to the target speaker. However, this approach increases the computational cost and is unsuitable for real-time processing. In this paper, we address both the speech over-suppression and interference leakage problems in causal PSE without increasing the inference cost.

\section{System Description}

\subsection{Baseline PSE and Problem}
\vspace{-.4em}

We build our PSE models based on the end-to-end enhancement network (E3Net) architecture of \cite{thakker2022fast} while the proposed approach is applicable to other model architectures. 
E3Net uses a learnable encoder and decoder. The encoded features are concatenated with a speaker embedding vector (d-vector) and fed into a stack of long short-term memory (LSTM) blocks. Each LSTM block consists of two fully connected layers, an LSTM layer with residual connection, and layer normalization modules.
On top of the last LSTM block, it has a fully connected layer for generating feature masks. 
They are multiplied with the encoded features and transformed into a waveform with the decoder to estimate the target speaker audio. 
The model is trained to minimize a power-law compressed phase-aware (PLCPA) loss function, which is defined as~\cite{scPSE}:
\begin{equation}\label{loss}
\begin{split}
&\mathcal{L}_{S,\hat{S}}(t, f) =  \alpha~\left||S(t, f)|^p-|\hat{S}(t, f)|^p\right|^2 + \\ &~~~~~~~(1-\alpha)~\left||S(t, f)|^pe^{j\varphi (S(t, f))}-|\hat{S}(t, f)|^pe^{j\varphi (\hat{S}(t, f))}\right|^2,
\end{split}
\end{equation}
where $\hat{S}(t,f)$ and $S(t,f)$ are the estimated and clean speech signals, respectively, at time $t$ and frequency $f$ in the short-time Fourier transform domain. Operator $\varphi$ calculates the argument of a complex number. The loss to be minimized is obtained by averaging $\mathcal{L}_{S,\hat{S}}$ over all time and frequency units. See \cite{thakker2022fast} for further details.

E3Net is causal and was shown to achieve good accuracy with a low computational cost. In \cite{thakker2022fast}, as with other prior models, the E3Net model was trained by using a dataset that always contained target speech signals. However, this training scheme promotes an interference leakage behavior when the input signal does not contain the target speaker at all or for a long time.

\subsection{PSE Training with Cross-task Knowledge Distillation}
\vspace{-.4em}

A naive approach to address the interference leakage issue would be to include ITS samples during training. An ITS sample is a noisy signal where the target speaker corresponding to the provided d-vector is completely inactive or silent. For the ITS samples, a PSE model is supposed to generate zero signals. Training the PSE model with the ITS samples helps it learn to remove the interfering speech signals when the target speaker is inactive. However, this simple approach tends to make the trained model so aggressive that it frequently attenuates the target speaker's speech signal. 
 
% Our hypothesis about the mechanism with which using the ITS samples during training results in an increase in speech over-suppression is as follows. 
Our hypothesis about the root cause of the increased speech over-suppression when the ITS samples are used during training is as follows.
% Due to the causality constraint and the limited model capacity for real-time operation, there will be interference-only time frames that are difficult to identify as not including the target speech. 
Due to the causality constraint and the limited model capacity for real-time operation, there will be some frames that are difficult to identify as target speech or interference. 
Forcing the PSE model to generate zero signals in these frames will encourage the model to occasionally zero out the signal gain even when the target speaker is present, worsening the speech over-suppression problem. 

To circumvent this problem and break the trade-off between the speech over-suppression and the interference leakage, we propose a cross-task distillation approach. 
Specifically, we use a separately trained causal personalized voice activity detection (pVAD) model to detect the challenging frames in the ITS training samples and exclude them from the PSE loss calculation. 
The pVAD model performs two-way classification for each frame to produce the posterior probability of each frame being spoken by the target speaker or not. Our pVAD model is based on a modified E3Net architecture. Instead of a learnable encoder, the pVAD model adopts 40-dimensional log Mel-filterbank energies as input by using the same window and hope size as the E3Net PSE model. The masking and decoder layers are replaced with a softmax layer for classification. 
The interference-only time frames of the ITS samples that the pVAD model misclassifies as containing the target speech are regarded as the challenging frames and removed from the PSE loss computation either with hard or soft decisions, as described below.

We examine three methods to modify the PSE loss for handling the misclassified frames by pVAD. In the first method, we explicitly exclude the misclassified frames from the PSE loss:
\begin{equation}\label{loss_frame_exclusion}
\mathcal{L}^{\tau}_{E}(t, f) =  c_\tau(t)\mathcal{L}_{S,\hat{S}}(t, f), 
\end{equation}
where
\begin{equation}\label{frame_exclusion}
\begin{split}
c_\tau(t) &= \begin{cases}
    1 & \text{if $p_{ts}(t) < \tau$}\\
    0 & \text{otherwise}
  \end{cases}, \\
\end{split}
\end{equation}
with $p_{ts}$ and $\tau$ being the frame-wise target speaker posterior probability generated by the pVAD model and a threshold, respectively. Fig.~\ref{fig_e3net_pvad} depicts the diagram of the proposed loss functions. For frame $t$, the target speaker is considered active by the pVAD model when $c_\tau(t)=0$ and inactive when $c_\tau(t)=1$. We use Eq.~\eqref{loss_frame_exclusion} as the PSE loss function to exclude the contribution of the misclassified frames in the ITS samples. Alternatively, the second method leaks the misclassified frames by using the noisy signal $Y$ as the reference signal: 
\begin{equation}\label{loss_mixture_as_ref}
\begin{split}
\mathcal{L}^{\tau}_{M}(t,f) &= \begin{cases}
    \mathcal{L}_{S,\hat{S}}(t,f) & \text{if $c_\tau(t) =1$}\\
    \mathcal{L}_{Y,\hat{S}}(t,f) & \text{otherwise}
  \end{cases}. \\
\end{split}
\end{equation}
This method alleviates target speaker over-suppression at the cost of slightly increased interference leakage. In the third method, we adjust the contribution of each frame by using the active target speaker probability as a weight in the PSE loss calculation: 
 \begin{equation}\label{loss_posterior_prob_as_weight}
\mathcal{L}_{P}(t, f) =  (1-p_{ts}(t))\mathcal{L}_{S,\hat{S}}(t, f).
\end{equation}
Intuitively, Eq.~\eqref{loss_posterior_prob_as_weight} reduces the loss contribution from misclassified frames and instead emphasizes the frames that are correctly predicted as an inactive target speaker. Unlike the previous methods, Eq.~\eqref{loss_posterior_prob_as_weight} does not require the threshold. Note that the proposed cross-task knowledge distillation scheme is applied only to the ITS training samples and that we use Eq.~\eqref{loss} as the loss function for the training samples containing the target speech.

\begin{table*}[!t]
	\centering
\caption{Comparison of different PSE training methods for TS1, TS2, and TS3 scenarios. TS1 includes the target, interfering speaker, and noise, while TS2 includes the target speaker and noise. TS3 includes only interfering speakers and noise. System U1 means unprocessed audio, B$\ast$ are baseline PNS models trained with Eq. \eqref{loss}, and S$\ast$ systems are based on the proposed training methods. All models used E3Net.}
\vspace{-.7em}
%For WER, DEL, and TSOS metrics, the lower is better. For DNSMOS, $\Delta N$ and STOI metrics, the higher is better. 
\label{tab:TS1_TS2}
	\renewcommand{\arraystretch}{1.00}
  \resizebox{\textwidth}{!}{
	  \begin{tabular}{{c|c | c | ccccc | ccccc | c}}
	  \toprule
	  
	    \multirow{2}{*}{System} & Train & \multirow{2}{*}{Loss} & \multicolumn{5}{c|}{TS1} & \multicolumn{5}{c|}{TS2} & \multicolumn{1}{c}{TS3} \\
         & data &  &WER$\downarrow$ &DEL$\downarrow$ &DNSMOS$\uparrow$ &STOI$\uparrow$ & TSOS$\downarrow$  &  WER$\downarrow$ & DEL$\downarrow$ & DNSMOS$\uparrow$ & STOI$\uparrow$ & TSOS$\downarrow$  & $\Delta N$$\uparrow$ \\
        
        \midrule
        U1 & -- &--& 43.0&	3.99&	2.92&	78.9&	0.00&		13.4&	2.00&	2.98&	85.0&	0.00 &	0.0 \\
        \midrule
        B1 &Base &$\mathcal{L}_{S,\hat{S}}$&31.9&    4.56&    3.56&    88.8&    1.35&        16.8&    2.55&    3.80&    93.4&    0.45&    46.5\\[2pt]
%        \cmidrule{3-14}
        B2 & Base/ITS &$\mathcal{L}_{S,\hat{S}}$&35.9&    8.28&    3.49&    85.5&    3.95&        20.3&    6.17&    3.70&    89.7&    2.54&    148.3\\
        \midrule
        S1 &  &$\mathcal{L}^{0.5}_{E}$&34.7&    4.46&    3.52&    88.6&    1.66&        17.8&    2.53&    3.75&    93.3&    0.37&    148.5\\[2pt]
%        \cmidrule{3-14}
        S2 &  &$\mathcal{L}^{0.25}_{E}$&34.5&    4.65&    3.53&    88.6&    1.98&        17.2&    2.68&    3.76&    93.2&    0.72&    148.4\\[2pt]
%        \cmidrule{3-14}
        S3 & Base/ITS &$\mathcal{L}^{0.1}_{E}$&35.1&    4.90&    3.50&    88.2&    2.06&        17.8&    3.01&    3.74&    92.8&    1.15&    148.3\\[2pt]
%        \cmidrule{3-14}
        S4 &  &$\mathcal{L}_{P}$&35.7&    5.73&    3.50&    87.5&    1.80&       18.4&    3.48&    3.72&    92.4&    0.48&    148.6\\[2pt]
%        \cmidrule{3-14}
        S5 & &$\mathcal{L}^{0.5}_{M}$&35.6&	4.73&	3.48&	87.8&	1.19&		18.1&	2.65&	3.70&	92.9&	0.32&	145.6\\
 \bottomrule
	  \end{tabular}
  }
\end{table*}

\begin{table*}[!t]
	\centering
\caption{Experimental results with asymmetric loss function.}
\vspace{-.7em}
%For WER, DEL, and TSOS metrics, the lower is better. For DNSMOS, $\Delta N$ and STOI metrics, the higher is better. 
\label{tab:TS1_TS2_ASYM}
	\renewcommand{\arraystretch}{1.00}
  \resizebox{\textwidth}{!}{
	  \begin{tabular}{{c|c | c | ccccc | ccccc | c}}
	  \toprule
	  
	    \multirow{2}{*}{System}& Train & \multirow{2}{*}{Loss} & \multicolumn{5}{c|}{TS1} & \multicolumn{5}{c|}{TS2} & \multicolumn{1}{c}{TS3} \\
        & data &  &WER$\downarrow$ &DEL$\downarrow$ &DNSMOS$\uparrow$ &STOI$\uparrow$ & TSOS$\downarrow$  &  WER$\downarrow$ & DEL$\downarrow$ & DNSMOS$\uparrow$ & STOI$\uparrow$ & TSOS$\downarrow$  & $\Delta N$$\uparrow$ \\
        \midrule
        B1a & Base &$\mathcal{L}_{S,\hat{S}}+\mathcal{L_{OS}}$&32.3&    3.70&    3.59&    90.3&    0.13&       16.5&    2.28&    3.79&    94.0&    0.03&    32.4\\[2pt]
        B2a & Base/ITS &$\mathcal{L}_{S,\hat{S}}+\mathcal{L_{OS}}$&35.7&    4.24&    3.55&    89.3&    0.90&       18.1&    2.73&    3.75&    93.4&    0.56&    145.6\\[2pt]
        S4a & Base/ITS &$\mathcal{L}_{P}+\mathcal{L_{OS}}$ &34.7&    5.56&    3.46&    87.1&    1.81&       17.9&    2.60&    3.70&    93.1&    0.27&    148.6\\

 \bottomrule
	  \end{tabular}
  }
\end{table*}

\section{Experimental Results}

We conducted a comprehensive test for the proposed training method by using datasets covering various conditions and performance metrics to evaluate different aspects of PSE systems. 

\subsection{Datasets}\label{sec_dataset}
\vspace{-.4em}
The evaluation was carried out based on simulated datasets. 
Room impulse responses (RIRs) were generated by using the image method with reverberation time (T60) between 0.15 and 0.6 seconds. In our simulation, we assumed  the target speaker to be closer to the microphone than the interfering speaker, which seems a reasonable assumption for telecommunication applications. The target speaker's distance to the microphone was in the range of (0, 1.3] m, while the interfering speaker's distance was greater than 2 meters.

We generated 2,000 and 50 hours of audio for the training and validation datasets, respectively, based on the clean speech data of the Deep Noise Suppression challange~\cite{reddy2021interspeech}. The clean speech signals were corrupted by the simulated RIRs and the noises from the AudioSet and Freesound datasets \cite{gemmeke2017audio,fonseca2017freesound} with signal-to-noise ratios (SNRs) in the range of [0, 15] dB. Half of the training and validation utterances contained the target and interfering speakers as well as noise with signal-to-interference ratios (SIRs) between 0 and 10 dB. The other half contained samples comprising the target speaker and noise only. The sampling rate was 16 kHz. The d-vectors had 128 dimensions and were extracted with a pre-trained Res2Net model (see~\cite{zhou2021resnext} for the details).
For training PNS models with ITS samples, we also randomly replaced the clean target speech with a zero signal in 15\% of the above training data to simulate an inactive target speaker scenario. We refer to the training datasets without and with the ITS samples as Base and Base/ITS, respectively.

The voice cloning toolkit (VCTK) corpus was used to create the test sets. The VCTK dataset contains clean utterances of 109 speakers with different English accents. We set aside 30 utterances of each speaker for d-vector extraction. To simulate a teleconference session, we concatenated the noisy reverberant mixtures generated from the same speaker's utterances to create a single long audio file for each speaker. The average audio file duration  was 27.5 minutes. 
The following three test sets were created to evaluate PSE models in different scenarios. TS1: the target speech signal is corrupted by both interfering speech and background noise; TS2: the target signal is corrupted by background noise; and TS3: the target speaker is inactive for the whole session and the audio file includes only interfering speakers and noise.
%TS3: there is only the target speaker (with reverberation;

\subsection{Evaluation Metrics}
\vspace{-.4em}
We used the word error rate (WER), deletion error rate (DEL), short-time objective intelligibility (STOI)~\cite{taal2011algorithm}, and DNSMOS~\cite{gamper2019intrusive} for performance measurement. DNSMOS is a neural network-based mean opinion score (MOS) estimator which was shown to be highly correlated with subjective quality ratings.
To directly measure the target speech over-suppression (TSOS) at the signal level, in addition to DEL, we also used the TSOS metric proposed in ~\cite{scPSE}. For each time frame, it is defined as 
\begin{equation}\label{tsos}
\begin{split}
\mathcal{TSOS}(t) &= \begin{cases}
    1 & \text{if $\sum_{f} \mathcal{L_{OS}}(t, f) > \gamma \sum_{f} |S(t, f)|^p$}\\
    0 & \text{otherwise}
  \end{cases},
% \mathcal{TSOS}(t) &= \sum_{f} \mathcal{L_{OS}}(t, f) > \gamma \sum_{f} |S(t, f)|^p
\end{split}
\end{equation}
where $\mathcal{L_{OS}}$ represents the following over-suppression index:
\begin{equation}\label{asym_loss}
\mathcal{L_{OS}}(t, f) =\left|\text{ReLU}(|S(t, f)|^p-|\hat{S}(t, f)|^p)\right|^2.
\end{equation}
$\text{ReLU}(.)$ is the rectified linear unit function, and $\gamma$ is a threshold value set at 0.1. Note that Eq.~\eqref{asym_loss} is a special version of the asymmetric loss proposed in~\cite{Wang2020}. 
%TSOS measures the target speaker over-suppression at the signal level. 
Since reference clean utterances occasionally contained modest non-speech sounds, we applied forced alignment to ignore the time frames with no speech activity. To make it easy to interpret the resultant numbers, we counted the segments where the frame-level TSOS values continued to be one for one second or longer.
%and normalized the segment count to a per-hour basis. 
Finally, to measure the interference leakage in the TS3 scenario, we calculated the energy difference between the input and residual signals, i.e., 
\begin{equation}\label{noise_diff}
\Delta N = 10\log|Y|^2 - 10\log|\hat{S}|^2. 
\end{equation}

\subsection{Implementation Details}
\vspace{-.4em}
Following~\cite{thakker2022fast}, we used an E3Net model consisting of 4 LSTM blocks and an encoder-decoder pair with 2,048 filters. The dimensions of the LSTM and fully connected layers of each LSTM block were 256 and 1024, respectively. We set the window size to 20 ms and the hop size to 10 ms. During the training, we generated mixtures on the fly by randomly selecting reverberated target and interfering speech signals and noise samples. We also applied a signal-domain variant of SpecAugument~\cite{Park2019} to input mixtures. The PLCPA loss parameters were set as $p=0.3$ and $\alpha=0.5$. 
The value of threshold $\tau$ was set at  0.5 by default for $\mathcal{L}^{\tau}_{E}$ and $\mathcal{L}^{\tau}_{M}$ losses.

% Each E3Net model was trained for 500K steps with a cosine annealing learning rate scheduler and a peak learning rate of $10^{-3}$. The best checkpoint was selected according to the validation loss.

Our pVAD model, used for the cross-task knowledge distillation, was based on E3Net with 40-dimensional log mel-filterbank input, three LSTM blocks, and a two-way softmax output layer. The model was trained with a binary cross-entropy loss on the Base/ITS training dataset. The ground-truth target speaker activity labels were generated by applying a DNN-based VAD model~\cite{braun2021training} to the underlying clean speech signals.

\subsection{Results and Discussions}
\vspace{-.4em}
Table~\ref{tab:TS1_TS2} shows the experimental results. Two baseline E3Net models were built based on PLCPA loss $\mathcal{L}_{S,\hat{S}}$ without the proposed cross-task scheme. One was trained on the Base training dataset (B1), and the other used the Base/ITS dataset (B2).
We can observe that including the ITS samples during training significantly reduced interference leakage in the TS3 scenario.  The average noise energy of TS3 was 148.8 dB, which means that the B2 model removed the noise and interference signals almost completely when the target speaker was silent. However, B2 considerably increased speech over-suppression in TS1 and TS2 compared with B1. For example, DEL and TSOS were increased by 3.62 percentage points and 2.09 seconds, respectively, in the T2 scenario. This shows the previous training scheme using PLCPA loss suffers from the trade-off between the speech over-suppression and the interference leakage.

Model S1 trained with the proposed loss of $\mathcal{L}^{0.5}_{E}$ yielded the DEL and TSOS values that are close to the results of B1 for both TS1 and TS2 while achieving almost the same $\Delta N$ value as B3 for TS3. 
This means that 
excluding the ITS frames misclassified by the pVAD model led to decreasing the interference leakage without incurring increases in speech over-suppression.
While S1 modestly increased the WER compared with B1, especially for the TS1 scenario (31.89\% $\to$ 34.67\%), which might be due to increased processing artifacts, the gain (46.52 dB $\to$ 148.49 dB) in TS3 was much more significant.

The effect of the threshold  in $\mathcal{L}^{\tau}_{E}$ was examined by 
changing the $\tau$ value to 0.25 and 0.1 (see S2 and S3). 
By decreasing $\tau$, more frames would be considered as being misclassified and would be excluded from the loss function. The results show that decreasing the pVAD threshold value did not lead to further speech over-suppression reduction. 
Instead of using hard pVAD decisions, 
training the PNS model with soft decisions using $\mathcal{L}_{P}$ loss resulted in a similar performance for all metrics (S5) without threshold adjustment. 
Finally, using $\mathcal{L}^{0.5}_{M}$ as the loss function further reduced the speech over-suppression measured by TSOS at the slight expense of the interference leakage. 
All results show that the proposed cross-task knowledge distillation training method based on pVAD improved the PNS performance in both speech over-suppression and leakage interference.

% asymmetric Loss
Table \ref{tab:TS1_TS2_ASYM} shows the results of the PNS models obtained by combining the asymmetric loss of \eqref{asym_loss} with PLCPA or the proposed loss. 
As we can see in B1a, this improved the baseline system with respect to the speech over-suppression at the cost of increased interference leakage in TS3. 
Adding the ITS training samples mitigated interference leakage significantly while showing a reasonable performance with respect to the speech over-suppression (B2a). 
The proposed method, denoted as S4a, further reduced the interference leakage amount while the TSOS results were mixed compared with B2a (i.e., improvement was observed for TS2 while the TSOS was degraded in TS1).

\section{Conclusion}
\vspace{-.4em}
In this work, we introduced a new causal PSE model training method to reduce interference leakage when the target speaker is inactive without over-suppressing the target speech. We used a pVAD task for cross-task knowledge distillation to achieve this goal. Specifically, we used misclassification patterns of a pVAD model to identify challenging frames of ITS training samples and excluded or de-emphasized them from the PNS model loss calculation. The experimental results showed the effectiveness of the proposed method.

\bibliographystyle{IEEEtran}
{\small\bibliography{refs}}

\begin{thebibliography}{10}
\providecommand{\url}[1]{#1}
\csname url@samestyle\endcsname
\providecommand{\newblock}{\relax}
\providecommand{\bibinfo}[2]{#2}
\providecommand{\BIBentrySTDinterwordspacing}{\spaceskip=0pt\relax}
\providecommand{\BIBentryALTinterwordstretchfactor}{4}
\providecommand{\BIBentryALTinterwordspacing}{\spaceskip=\fontdimen2\font plus
\BIBentryALTinterwordstretchfactor\fontdimen3\font minus
  \fontdimen4\font\relax}
\providecommand{\BIBforeignlanguage}[2]{{%
\expandafter\ifx\csname l@#1\endcsname\relax
\typeout{** WARNING: IEEEtran.bst: No hyphenation pattern has been}%
\typeout{** loaded for the language `#1'. Using the pattern for}%
\typeout{** the default language instead.}%
\else
\language=\csname l@#1\endcsname
\fi
#2}}
\providecommand{\BIBdecl}{\relax}
\BIBdecl

\bibitem{scPSE}
S.~E. Eskimez, T.~Yoshioka, H.~Wang, X.~Wang, Z.~Chen, and X.~Huang,
  ``Personalized speech enhancement: New models and comprehensive evaluation,''
  in \emph{Proc. ICASSP}, 2022, pp. 356--360.

\bibitem{giri2021personalized}
R.~Giri, S.~Venkataramani, J.-M. Valin, U.~Isik, and A.~Krishnaswamy,
  ``Personalized {PercepNet}: Real-time, low-complexity target voice separation
  and enhancement,'' in \emph{Proc. Interspeech}, 2021, pp. 1124--1128.

\bibitem{taherian2022one}
H.~Taherian, S.~E. Eskimez, T.~Yoshioka, H.~Wang, Z.~Chen, and X.~Huang, ``One
  model to enhance them all: Array geometry agnostic multi-channel personalized
  speech enhancement,'' in \emph{Proc. ICASSP}, 2022, pp. 271--275.

\bibitem{delcroix2020improving}
M.~Delcroix, T.~Ochiai, K.~Zmolikova, K.~Kinoshita, N.~Tawara, T.~Nakatani, and
  S.~Araki, ``Improving speaker discrimination of target speech extraction with
  time-domain speakerbeam,'' in \emph{Proc. ICASSP}, 2020, pp. 691--695.

\bibitem{Wang2020}
Q.~Wang, I.~L. Moreno, M.~Saglam, K.~Wilson, A.~Chiao, R.~Liu, Y.~He, W.~Li,
  J.~Pelecanos, M.~Nika, and A.~Gruenstein, ``{VoiceFilter-Lite}: Streaming
  targeted voice separation for on-device speech recognition,'' in \emph{Proc.
  Interspeech}, 2020, pp. 2677--2681.

\bibitem{delcroix2022listen}
M.~Delcroix, K.~Kinoshita, T.~Ochiai, K.~Zmolikova, H.~Sato, and T.~Nakatani,
  ``{Listen only to me! How well can target speech extraction handle false
  alarms?}'' in \emph{Proc. Interspeech}, 2022, pp. 216--220.

\bibitem{thakker2022fast}
M.~Thakker, S.~E. Eskimez, T.~Yoshioka, and H.~Wang, ``Fast real-time
  personalized speech enhancement: End-to-end enhancement network ({E3Net}) and
  knowledge distillation,'' in \emph{Proc. Interspeech}, 2022, pp. 991--995.

\bibitem{Chen2022MultiStage}
L.~Chen, C.~Xu, X.~Zhang, X.~Ren, X.~Zheng, C.~Zhang, L.~Guo, and B.~Yu,
  ``Multi-stage and multi-loss training for fullband non-personalized and
  personalized speech enhancement,'' in \emph{Proc. ICASSP}, 2022, pp.
  9296--9300.

\bibitem{9747765}
Y.~Ju, W.~Rao, X.~Yan, Y.~Fu, S.~Lv, L.~Cheng, Y.~Wang, L.~Xie, and S.~Shang,
  ``{TEA-PSE}: {Tencent-Ethereal-Audio-Lab} personalized speech enhancement
  system for {ICASSP} 2022 {DNS} challenge,'' in \emph{Proc. ICASSP}, 2022, pp.
  9291--9295.

\bibitem{borsdorf21}
M.~Borsdorf, C.~Xu, H.~Li, and T.~Schultz, ``Universal speaker extraction in
  the presence and absence of target speakers for speech of one and two
  talkers,'' in \emph{Proc. Interspeech}, 2021, pp. 1469--1473.

\bibitem{Zhang2020}
Z.~Zhang, B.~He, and Z.~Zhang, ``{X-TaSNet}: Robust and accurate time-domain
  speaker extraction network,'' in \emph{Proc. Interspeech}, 2020, pp.
  1421--1425.

\bibitem{reddy2021interspeech}
C.~K. Reddy, H.~Dubey, K.~Koishida, A.~Nair, V.~Gopal, R.~Cutler, S.~Braun,
  H.~Gamper, R.~Aichner, and S.~Srinivasan, ``{INTERSPEECH} 2021 deep noise
  suppression challenge,'' in \emph{Proc. Interspeech 2021}, 2021, pp.
  2796--2800.

\bibitem{gemmeke2017audio}
J.~F. Gemmeke, D.~P. Ellis, D.~Freedman, A.~Jansen, W.~Lawrence, R.~C. Moore,
  M.~Plakal, and M.~Ritter, ``Audio set: An ontology and human-labeled dataset
  for audio events,'' in \emph{Proc. ICASSP}, 2017, pp. 776--780.

\bibitem{fonseca2017freesound}
E.~Fonseca, J.~Pons~Puig, X.~Favory, F.~Font~Corbera, D.~Bogdanov, A.~Ferraro,
  S.~Oramas, A.~Porter, and X.~Serra, ``Freesound datasets: a platform for the
  creation of open audio datasets,'' in \emph{Proc. ISMIR}, 2017, pp. 486--493.

\bibitem{zhou2021resnext}
T.~Zhou, Y.~Zhao, and J.~Wu, ``{ResNeXt} and {Res2Net} structures for speaker
  verification,'' in \emph{Proc. SLT}, 2021, pp. 301--307.

\bibitem{taal2011algorithm}
C.~H. Taal, R.~C. Hendriks, R.~Heusdens, and J.~Jensen, ``An algorithm for
  intelligibility prediction of time--frequency weighted noisy speech,''
  \emph{IEEE Transactions on Audio, Speech, and Language Processing}, vol.~19,
  pp. 2125--2136, 2011.

\bibitem{gamper2019intrusive}
H.~Gamper, C.~Reddy, R.~Cutler, I.~Tashev, and J.~Gehrke, ``Intrusive and
  non-intrusive perceptual speech quality assessment using a convolutional
  neural network,'' in \emph{Proc. WASPAA}, 2019.

\bibitem{Park2019}
D.~S. Park, W.~Chan, Y.~Zhang, C.-C. Chiu, B.~Zoph, E.~D. Cubuk, and Q.~V. Le,
  ``{SpecAugment}: A simple data augmentation method for automatic speech
  recognition,'' in \emph{Proc. Interspeech}, 2019, pp. 2613--2617.

\bibitem{braun2021training}
S.~Braun and I.~Tashev, ``On training targets for noise-robust voice activity
  detection,'' in \emph{Proc. EUSIPCO}, 2021, pp. 421--425.

\end{thebibliography}

\end{document}